\documentclass[prl,twocolumn,aps,amsmath,amssymb,showpacs,superscriptaddress]{revtex4-1}

\usepackage{graphics}
\usepackage{dcolumn}
\usepackage{bm}
\usepackage[T1]{fontenc}

\begin{document}

\title{Shot Noise in Lithographically Patterned Graphene Nanoribbons}

\author{Z. B. Tan}
\affiliation{O.V. Lounasmaa Laboratory, Aalto University, P.O. Box 15100, FI-00076 AALTO, Finland}
\author{A. Puska}
\affiliation{O.V. Lounasmaa Laboratory, Aalto University, P.O. Box 15100, FI-00076 AALTO, Finland}
\author{T. Nieminen}
\affiliation{O.V. Lounasmaa Laboratory, Aalto University, P.O. Box 15100, FI-00076 AALTO, Finland}
\author{F. Duerr}
\affiliation{Physikalisches Institut (EP3), University of W\"urzburg, 97074 W\"urzburg, Germany}
\author{C. Gould}
\affiliation{Physikalisches Institut (EP3), University of W\"urzburg, 97074 W\"urzburg, Germany}
\author{L. W. Molenkamp}
\affiliation{Physikalisches Institut (EP3), University of W\"urzburg, 97074 W\"urzburg, Germany}
\author{B. Trauzettel}
\affiliation{Institute for Theoretical Physics and Astrophysics, University of W\"urzburg, W\"urzburg, Germany}
\author{P. J. Hakonen}\email[Corresponding author: pertti.hakonen@aalto.fi]{}
\affiliation{O.V. Lounasmaa Laboratory, Aalto University, P.O. Box 15100, FI-00076 AALTO, Finland}

\begin{abstract}

We have investigated shot noise and conductance of multi-terminal graphene nanoribbon devices at temperatures down to 50 mK. Away from the charge neutrality point, we find a Fano factor $F \approx 0.4$, nearly independent of the charge density.  Our shot noise results are consistent with theoretical models for disordered graphene ribbons with a dimensionless scattering strength $K_0 \approx 10$ corresponding to rather strong disorder. Close to charge neutrality, an increase in $F$ up to $\sim 0.7$ is found, which indicates the presence of a dominant Coulomb gap possibly due to a single quantum dot in the transport gap.

\end{abstract}


\maketitle


Electrical conduction in graphene ribbons is strongly influenced by disorder which brings about localization of charge carriers and transport via hopping conduction. It was noted already by Mott \cite{mott1968} that hopping conduction at low temperatures results from states whose energies are located in a narrow band near the Fermi level. As a consequence, competition between thermal excitation and the overlap integrals between localized states leads to variable range hopping (VRH), the Mott law, with a characteristic temperature dependence $\propto \exp{1/T^\frac{1}{1+d}}$ where $d$ is the dimension of the system. Under electron-electron interactions the Mott law is modified and a Coulomb gap may be formed  \cite{shklovskii1984}.

There is no Coulomb gap in good metals. However, it was shown by Altshuler and Aronov \cite{AA} that, in disordered metals, the density of states has a minimum around the Fermi energy. The depth of this minimum due to the electron-electron interactions increases with the amount of disorder. As the disorder grows sufficiently large,  electronic states become fully localized and the density of states vanishes at the Fermi level, i.e., a Coulomb gap is formed.

The gradual approach towards localization and Coulomb gap can be  probed in graphene nanoribbons (GNR) as a function of charge density induced by a gate. Interestingly, the zero-bias anomaly grows monotonically when approaching the charge neutrality point (CNP). Our results suggest a scenario where, first, a series of quantum dots is formed and transport occurs by tunneling between quantum dots. As the tunneling in VRH takes place in the optimal band around the Fermi level, the optimum gap is initially larger than the Coulomb gap, and transport is governed by the tunneling/cotunneling between the adjacent quantum dots. When charge density is lowered, the role of disorder and Coulomb interactions becomes even more important. Eventually, the Coulomb gap exceeds Mott's optimum band and single particle states near the Fermi level become strongly suppressed. This leads to enhanced suppression in electric transport and the activation laws display a larger gap value than would be expected from Anderson type of localization alone. At the same time, shot noise may be enhanced, reflecting a reduction in the interacting sections that limit the tunneling conduction.

The first studies of GNRs down to width $W \simeq  20$ nm \cite{chen2007,han2007} demonstrated the presence of a transport gap inversely proportional to the width and independent on the crystallographic orientation \cite{han2007}.  Similar transport gaps were observed for much smaller ribbon width in GNRs fabricated using sonication of intercalated graphite in solution, indicating smoother edges than the etched GNRs \cite{li2008}. The experiments performed on GNRs \cite{chen2007,han2007,ozyilmaz2007,han2009,oostinga2010, danneau2010} have clearly indicated variable range hopping while the role of the Coulomb gap has remained elusive \cite{sols2007,Droscher2011,Review}. The role of interactions in VRH has been investigated also in quantum Hall regime \cite{Bennaceur2012}.  Here, we demonstrate that, according to shot noise experiments on high-quality etched GNRs, a Coulomb gap is formed which leads to enhanced shot noise in the graphene ribbon. Both the shot noise and the $I-V$  characteristics measured in the gap region are consistent with the conclusion that the Coulomb gap originates from a single dominant quantum dot that limits the charge transport.
Furthermore, our results show nearly constant shot noise as a function of gate voltage away from the charge neutrality region,  which is in good agreement with the numerical simulations of Lewenkopf {\it et al.} \cite{lewenkopf2008} on disordered graphene ribbons.

The GNRs, patterned into four terminal cross geometry, were fabricated from micromechanically cleaved graphene deposited on a heavily \emph{p}-doped substrate with 300 nm SiO$_2$ layer. The graphene sheet was first connected using standard e-beam lithography followed by a Ti(2 nm)/Au(35 nm) bilayer deposition with lift-off in acetone.
A second lithography step allowed the patterning of the GNRs. The resist (PMMA) was used as mask in this step and GNRs were etched using an oxygen/argon plasma. We studied various lead configurations and found equivalent results on them. Here, we present only results on the configuration indicated in Fig. 1: measurement through the cross with the side terminals floating.  The length and width of the arms was $L \sim 240$ nm  and $W \sim 50$ nm, respectively, which yields a total length of $L=530$ nm. After the experiments, the GNRs were observed using scanning electron microscope (see Fig. 1). The measurements were performed on a dry Bluefors dilution refrigerator down to 50 mK using a noise spectrometer described in Ref. \onlinecite{danneau2008}. A tunnel junction was used for calibration of the shot noise and nonlinearities were taken into account within the gap regime \cite{wu2007}.

\begin{figure}[htbp]
\scalebox{0.60}{\includegraphics{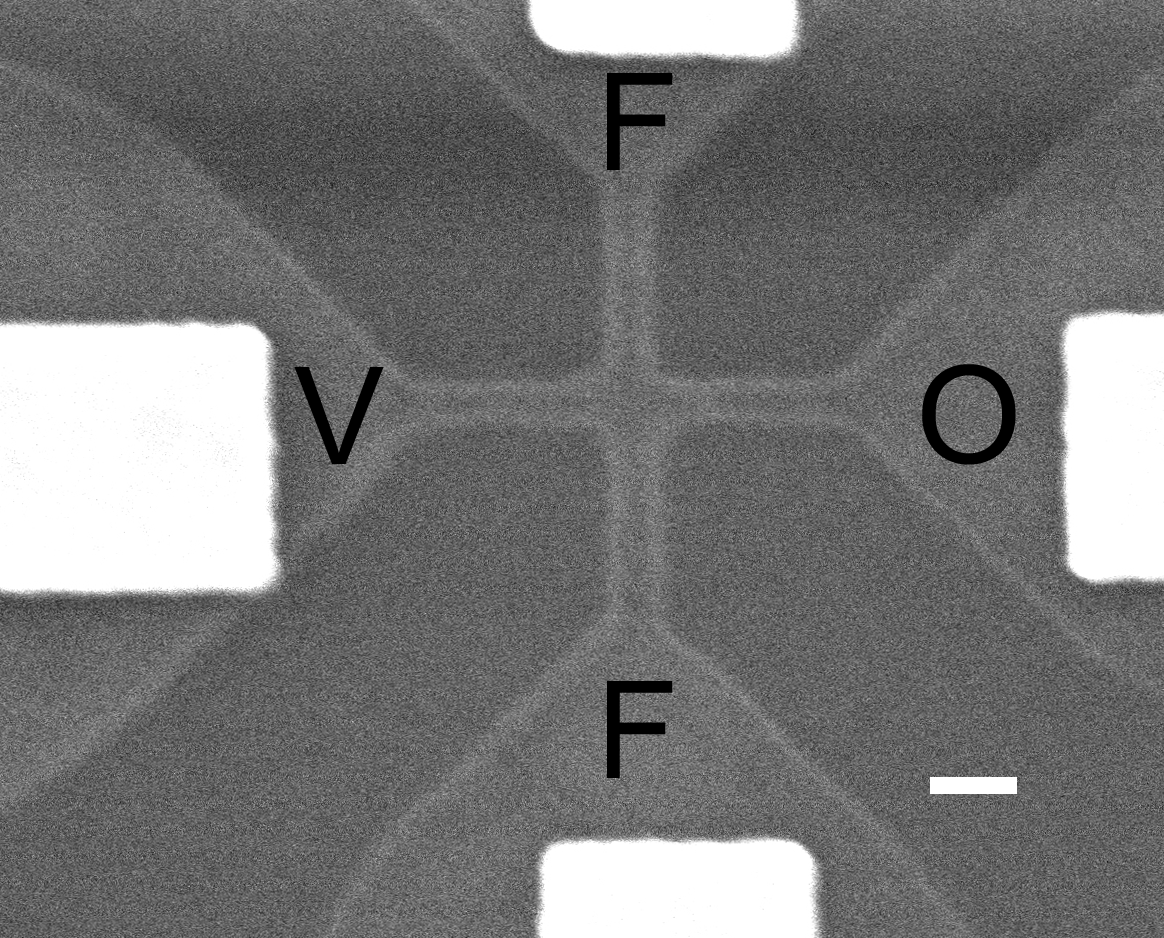}} {\caption{ Scanning electron micrograph of the graphene ribbon sample. Terminal V denotes the biased lead, while O is grounded and terminals F are floating. The white scale bar corresponds to 100 nm.}}
\end{figure}

Fig. 2a displays the gate voltage $V_{g}$ dependence of the differential conductance $G_d=dI/dV$ for zero bias voltage and for $V_b=100$ mV. In the zero bias data, there is a high impedance region at $V_g \simeq  3 ... +11$ V.
Clear conductance oscillations are visible, but no periodicity is detectable.
Far away from the charge neutrality point $G_d \sim 2e^2/h=g_0$, roughly equal to the conductance quantum $g_0$.
In Fig 2b, we display a color map of the logarithmic differential conductivity $\frac{W}{L}G_d$ on the bias plane spanned by $V_{b}$ and $V_{g}$. These findings coincide with the formation of a "large impedance region" or a "transport gap" as first observed in Refs.  \onlinecite{chen2007,han2007,ozyilmaz2007}.
The $I-V$ characteristics indicate a gap that is modulated by the diamond-like Coulomb structures which are typically assigned to the formation of a series of quantum dots. In Fig. 2b, the "drain source gap" amounts to $\sim 50$ meV.

\begin{figure}[htbp]
\scalebox{0.55}{\includegraphics{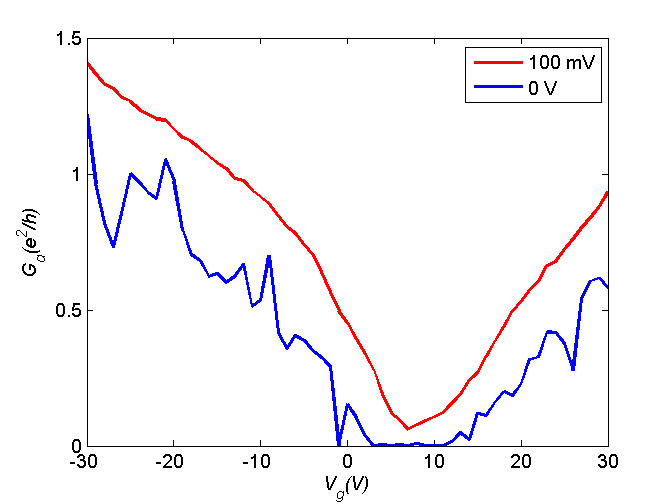}}

\scalebox{0.60}{\includegraphics{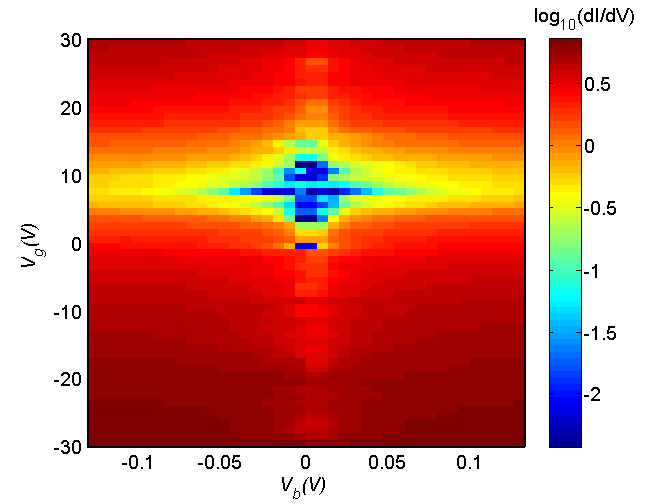}} {\caption{a) Differential conductance $G_d=\frac{dI}{dV}$ as a function of gate voltage $V_{g}$ at $V_b=0$ and ar $V_b = 100$ mV. b) Color map of logarithmic $g=\frac{W}{L} \frac{dI}{dV}$ versus bias voltage $V_{b}$ and gate voltage $V_{g}$ at $T =$ 50 mK. The charge neutrality point is located approximately at $V_g= 8$ V. }}
\end{figure}

Variable range hopping (VRH) generally describes electronic transport in the presence of disorder \cite{shklovskii1984}. Temperature dependence of the conductance $G(T)$ is conventionally used to identify the regime. In the case of graphene nanoribbons, however, $G(T)$ analysis is difficult to perform because the $V_{g}$ value of the minimum conductivity may change as the temperature is lowered  \cite {han2007}.  Consequently, we analyze $I-V$ curves at a fixed temperature $T$. Provided that there is a finite density of states at the Fermi level,
we may write for interaction dominated VRH:
\begin{eqnarray}
I(V,T) = VG_{0}(T) \exp \left\{-\left(\frac{V_0}{V} \right)^{1/2} \right\} \label{equ1} ,
\end{eqnarray}
which is valid up to voltages of $V_0$. In this expression, we do not write the dimensionality dependence of the exponent as we consider interaction dominated conduction at reduced dimension ($d= 1$ or 2) which lead to equivalent behavior. Eq.~\ref{equ1} transforms to Mott's law by replacement of $e V_0= k_B T_0$ and $e V = k_B T$ in the exponent ($V_0$ being the largest value for which the formula is valid) \cite{pollak,likharev,fogler}.

\begin{figure}[htbp]\label{V0}
\scalebox{0.55}{\includegraphics{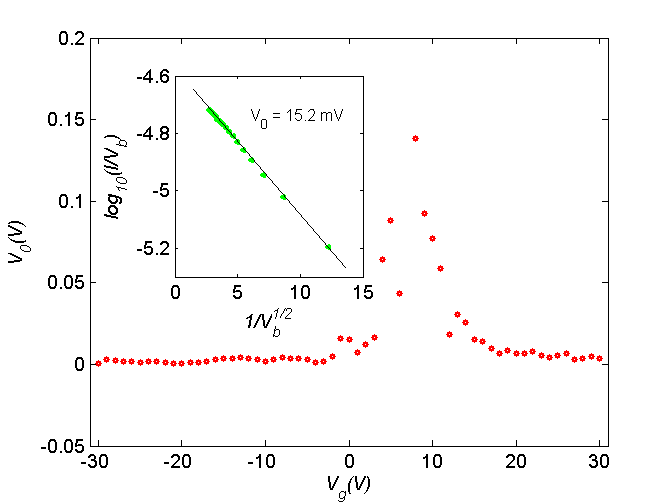}} {\caption{Inset: Measured $IV$ characteristics in terms of $\log(I/V)$ vs. $V^{-1/2}$ at $V_g=0$ V. The fitted line correspond to form of Eq. \ref{equ1}. The main frame illustrates the parameter $V_0$ as a function of $V_g$. }}
\end{figure}

The inset of Fig. 3 displays an $I-V$ curve for our sample measured at low bias. Comparing well with Eq.~(\ref{equ1}), we see that the conduction data follow nicely a VRH-like law over the gap region. The data in Fig. 3 are plotted as $\log(I/V)$ vs. $V^{-1/2}$, in accordance with the Coulomb interaction dominated transport at reduced dimensions. The exponential factor in the Coulomb interaction law is given by
${V_0} = \frac{\beta e^2}{\epsilon_0 \kappa \xi}$,
where the numerical factor $\beta \simeq 3$ \cite{shklovskii1984},  $\xi$ denotes the extent of the localized state, and $\kappa$ is the relative permittivity of the substrate. From Fig. 3, which displays $V
_0$ as a function of $V_g$, we find $V_0$ values up to $\sim 0.15$ V. The maximum value corresponds to the self-energy of an island on the order of size $\xi \simeq 40$ nm in a medium with effective $\kappa \sim 8$ \cite{Lian2010,Duerr2012}. This result is in accordance with experiments on similar graphene devices \cite{Review}. Assuming a uniform disorder, this means that we have $10-15$ islands of localized states  over our sample length.
However, uniform distribution of island dimensions is not in accordance with our noise data as will be discussed below.
\begin{figure}[htbp]
\scalebox{0.57}{\includegraphics{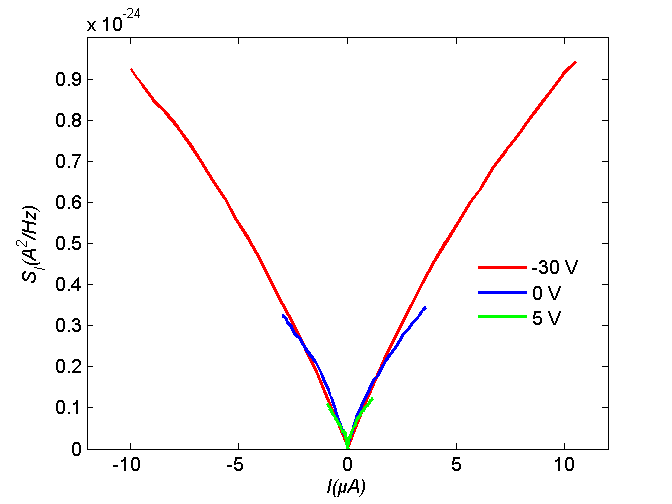}} {\caption{Shot noise $S_I$ vs. current at $V_g=$ -30, 0, and 5 V, respectively.}}
\end{figure}
\begin{figure}[htbp]
\scalebox{0.60}{\includegraphics{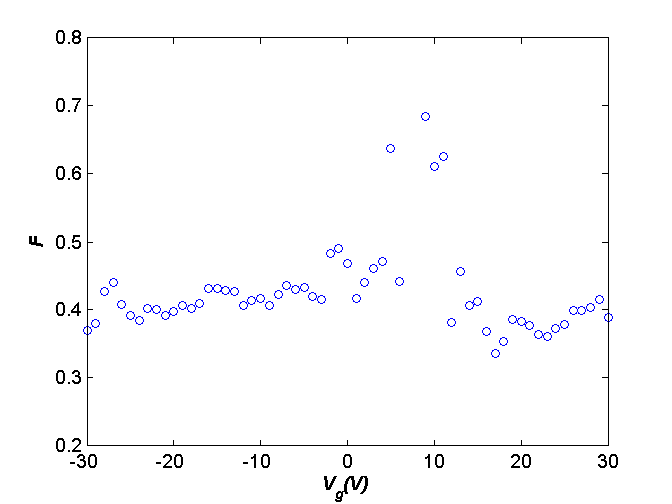}} {\caption{Low-bias Fano factor as a function of gate voltage.
}}
\end{figure}

In order to understand the role of disorder in graphene ribbons better, we have investigated shot noise at low temperatures down to 50 mK where the contribution of inelastic scattering events should be small and scattering matrix theory should be applicable. Shot noise denotes current fluctuations arising from the granular nature of the charge carriers (see Ref. \onlinecite{blanter2000} for a review). The Fano factor $F$, given by the ratio of shot noise and mean current, is commonly employed to quantify shot noise. The noise power spectrum then reads $S(I) = F \times 2eI$. In the scattering matrix formalism \cite{blanter2000}, $F = \sum_{n=1}^{N}T_n(1-T_n)/\sum_{n=1}^{N} T_n$, depending only on the transmission $T_n$ of the $N^{th}$
quantum channels. For diffusive conductors with a bimodal distribution of transmission eigenvalues we have $F=\frac{1}{3}$.
The case of graphene is unique since transport at the Dirac point occurs via evanescent waves and this gives rise to a Fano factor of $\frac{1}{3}$ for large width over length ratio ($\frac{W}{L}>3$) \cite{katsnelson2006a,danneau2008,dicarlo2008}. According to theory, smooth potential disorder tends to decrease $F$ \cite{sanjose2007}. However, when the disorder is strong even increased levels of noise can be observed \cite{lewenkopf2008}.

We have performed our shot noise measurements over the frequency range $f_{BW}=600-900$ MHz. This frequency is high enough so that typically all noise due to fluctuations of resistance (transmission coefficients) can be neglected. Nevertheless, $f_{BW}$ corresponds to the zero-frequency noise as it is low compared with the internal $RC$ time scales.

Our results on the current noise per unit bandwidth $S_I$  with increasing bias current are displayed in Fig. 4.  The data show linear slope with current at small bias, which bends weakly down at large voltages where inelastic phonon scattering events start to take place. For phase coherent transport at low bias, the shot noise can be described by scattering matrix theory, while in the incoherent regime at high bias, the noise can be understood using semiclassical models, and even a separation of noise contributions from different sources within the graphene sample can be made. When inelastic processes are important (inelastic length $l_{in} \lesssim L$, shot noise starts to decrease and is dependent on the details of the relaxation processes, acoustic or optical phonons \cite{danneau2008,fay2011} that govern the ensuing non-equilibrium state.

The initial slope of the traces in Fig. 4, i.e. the Fano factor $F$, is illustrated in Fig. 5 as a function of $V_g$.  In the regime of linear $IV$ curves, the $S_I$  \emph{vs}. $V$ curves were fitted using the formula defined in Ref. \onlinecite{danneau2008} with $F$ as the only fitting parameter.  Away from the CNP, the value of $F\simeq 0.4$ is in agreement with disordered graphene ribbons \cite{lewenkopf2008}. We conclude on the basis of Ref. \onlinecite{lewenkopf2008} that our noise measurement is in accordance with a Gaussian disorder corresponding to a dimensionless disorder strength of $K_0 \approx 10$ meaning that our sample is strongly affected by disorder.

Near the CNP, the conductivity data indicate transport via a series of quantum dots. The influence of electron-electron interactions on mesoscopic conductors has been considered in Refs. \onlinecite{galaktionov2003, golubev2004,golubev2005} which indicate that their effect on the Fano factor can be either positive or negative, depending on the magnitude of the transmission coefficient. Golubev and Zaikin have derived for the shot noise of an array of $N - 1$ interacting chaotic  quantum dots ($N$ barriers) \cite{golubev2004}
\begin{equation}
F  = \frac{1}{3} + \sum\limits_n^N {\frac{{R_n^3}}{{R_\Sigma ^3}}} \left( {{F _n} - \frac{1}{3}} \right)
\end{equation}
where $F_n$ and $R_n$ denote the Fano factor and resistance of the nth individual barrier and $R_{\Sigma} = \sum\limits_n^N {R_n}$. For scatterers with $F_n \neq 1/3$, the above equation yields about $1/3$ for a large number of $N$, indicating that for nearly uniform, long arrays of quantum dots -- even with $F_n \sim1$ -- we expect to have $F \simeq 1/3$. Indeed, this behavior is observed in our experiments around the onset of the gap regime, although the value of $F$ points towards a rather small number of islands. Our data indicate transport in the localized regime without any Fano factor suppression by electron-phonon scattering as was found in Ref. \onlinecite{danneau2010} at 4.2 K.

Super-Poissonian noise is possible for a series of quantum dots \cite{aghassi2006}. Typically, the strongly enhanced shot noise is related with switching between fast and slow transport modes which yields a Lorentzian spectrum, where the cut-off varies with bias \cite{wu2007b}. Super-Poissonian behavior has also been found in correlated resonant tunneling involving two interacting localized states \cite{safonov2003} and in carbon nanotubes \cite{onac2006}. However, our  value for $F$ near the CNP fits best the noise processes present in a single quantum dot. An asymmetric quantum dot will yield $F=0.5 - 1$, depending on the ratio of its tunneling barrier transparencies. Hence, we conclude that the noise is dominated by a single quantum dot and that the gap originates from the Coulomb gap produced by this single island.

To conclude, we have measured shot noise and conductance in graphene nanoribbons. Our results indicate that away from the charge neutrality point disorder is strongly affecting our transport data. This finding is most convincingly demonstrated by a nearly density independent Fano factor $F \approx 0.4$. Close to charge neutrality, different physics seems to be valid because Coulomb interactions start to become important. Then, the Fano factor increases, a behavior that is consistent with transport through a dominant quantum dot in the ribbon.

We thank D. Golubev, T. Heikkil\"a, M. Laakso, A. Zaikin, and F. Libisch for fruitful discussions. This work has been supported in part by the EU 7th Framework Programme (Grant No. 228464 Microkelvin and No. EU-FP7-NMP-246026), by the Academy of Finland (projects no. 250280 LTQ CoE and 132377), and by the European Science Foundation (ESF) under the EUROCORES Programme EuroGRAPHENE.

\end{document}